# Spin-transfer-torque efficiency enhanced by edge-damage of perpendicular magnetic random access memories

Kyungmi Song[1] and Kyung-Jin Lee[1,2,*]

[1]KU-KIST Graduate School of Converging Science and Technology, Korea University, Seoul 136-713, Korea

[2]Department of Materials Science and Engineering, Korea University, Seoul 136-713, Korea

* Corresponding email: kj_lee@korea.ac.kr

**We numerically investigate the effect of magnetic and electrical damages at the edge of a perpendicular magnetic random access memory (MRAM) cell on the spin-transfer-torque (STT) efficiency that is defined by the ratio of thermal stability factor to switching current. We find that the switching mode of an edge-damaged cell is different from that of an undamaged cell, which results in a sizable reduction in the switching current. Together with a marginal reduction of the thermal stability factor of an edge-damaged cell, this feature makes the STT efficiency large. Our results suggest that a precise edge control is viable for the optimization of STT-MRAM.**

## I.     INTRODUCTION

Traditional charge-based memory technologies are approaching miniaturization limits as it becomes increasingly difficult to reliably retain sufficient electrons in shrinking cells [1]. Alternative emerging memory technologies using resistance change rather than charge storage include STT-MRAM [2-8], phase-change RAM [9], and resistive RAM [10,11]. Among these nonvolatile memories, STT-MRAMs have attracted considerable attention owing to their excellent endurance and low power consumption.

Commercialization of STT-MRAMs requires a small switching current $I_{SW}$ and a large thermal stability factor $\Delta$ (= $E_B/k_BT$ where $E_B$ is the energy barrier and $k_BT$ is the thermal energy), which is parameterized by the STT-efficiency $\eta_{STT}$ (= $\Delta/I_{SW}$) [6]. Under the assumption of uniform magnetic properties across the MRAM cell and the single domain switching, $E_B$ and $I_{SW}$ are given by $K_{u,eff}V$ and $\alpha(2e/\hbar) K_{u,eff}V/P$, respectively, where the effective perpendicular anisotropy energy density $K_{u,eff} = K_u - \frac{1}{2}N_dM_S^2$, $N_d$ is the demagnetization factor, $V$ is the free-layer volume, $P$ is the effective spin polarization. As $K_{u,eff}V$ is common in both $E_B$ and $I_{SW}$, $\eta_{STT}$ is given by $P\hbar k_BT /(2e\alpha)$ that is fixed for a fixed layer structure and material choice. Because of the common factor in $E_B$ and $I_{SW}$, decoupling $E_B$ from $I_{SW}$ (i.e., a large $E_B$ as well as a small $I_{SW}$) is challenging fundamentally and technologically, which is a key obstacle for the commercialization of STT-MRAMs.

In this work, we show that this obstacle could be removed at least partially when the magnetic properties are properly modified at the edge of MRAM cell. We note that it is highly probable that the cell edges are damaged physically and/or chemically during etching process. An example of edge-damage in MRAM cells is that a resistance-area product RA measured by current-in-plane tunneling (CIPT) on unprocessed stacks is different from RA on patterned MRAM cell [12]. It was also reported that the exchange-bias of IrMn/CoFe is substantially damaged by ion milling process [13]. In this respect, it is important to investigate effects of possible edge-damage on the switching current and the energy barrier of perpendicular MRAM cells.

## II. MODEL

We calculate the energy barrier of a circular shaped perpendicular nanomagnet based on the Nudged Elastic Band (NEB) method [14-16]. To compute the energy minimum path between two states, the initial path of the switching is first assumed (generally the most probable path is guessed). Then the energy minimum path is obtained by minimizing the gradient of the energy [14-16]. We also compute the switching current by solving the Landau-Lifshitz-Gilbert (LLG) equations with the STT term as

$$\frac{d\hat{m}}{dt} = -\gamma \hat{m} \times \vec{H}_{eff} + \alpha \hat{m} \times \frac{d\hat{m}}{dt} + \gamma \frac{\hbar}{2e} \frac{\eta}{M_s t} J\, \hat{m} \times (\hat{m} \times \hat{p}), \qquad (1)$$

where the $\hat{m}$ is the unit vector along the magnetization, $\vec{H}_{eff}$ is the effective field including the exchange, magnetostatic, anisotropy and current-induced Oersted fields, $\alpha$ (= 0.01) is the damping constant, $\hat{p}$ (= $\hat{z}$) is the unit vector along the spin orientation of incoming spin current, $M_S$ is the saturation magnetization, $t$ (= 1 nm) is thickness, $\eta$ (= 0.5) is the spin polarization factor, and $J$ is the current density. The cell size is fixed to 1 x 1 x 1 nm$^3$ for the free layer. The fixed layer is not considered so that no stray field from the fixed layer is included. The time step for integration is 0.05 ps and all calculations are performed at T=0K.

We use following parameters for an undamaged cell: the perpendicular magnetic anisotropy energy density $K_u$ is $10^7$ erg/cm$^3$, $M_S$ is 1000 emu/cm$^3$, and the exchange stiffness constant $A_{ex}$ is $10^{-6}$ erg/cm. We model the edge-damaged region around the rim with the width of 5 nm (see Fig. 1). In edge-damaged cells, we assume that at least one of three magnetic properties ($K_u$, $M_S$, and $A_{ex}$) is degraded. We assume a radially linear degradation of each property in the edge-damaged region as follows: $K_u$ varies from $10^7$ to $10^6$ erg/cm$^3$, $M_s$ varies from 1000 to 500 emu/cm$^3$, and $A_{ex}$ varies from $10^{-6}$ to $10^{-7}$ erg/cm. We test seven cases, one undamaged case and six damaged cases such as 1) $K_u$ only, 2) $M_S$ only, 3) $A_{ex}$ only, 4) both $K_u$ and $M_S$, 5) both $A_{ex}$ and $M_S$, 6) all of $K_u$, $M_S$, and $A_{ex}$.

## III. RESULTS AND DISCUSSIONS

Figure 2 shows simulation results of (a) thermal stability factor $\Delta$ at $T$ = 300 K, (b) switching current density, and (c) corresponding STT-efficiency at various cell sizes. The edge-damage effect on the thermal stability factor is rather easy to understand by estimating the average effective perpendicular anisotropy energy density $<K_{u,eff}>$ where $<...>$ stands for the spatial average. In most edge-damaged cases, the thermal stability factor is reduced compared to the undamaged case (solid line). An exception is the $M_S$-damaged case (open circles) where the thermal stability factor is larger than that of the undamaged case for all tested cell sizes. It is because the reduced $M_S$ at the edge increases the overall effective perpendicular anisotropy energy density $<K_{u,eff}>$ through a reduced demagnetizing effect.

On the other hand, the switching current density exhibits more complicated dependence on the type of edge-damage. In comparison to the undamaged case, three edge-damaged cases, $K_u$-damaged (solid upper triangles), $M_S$-$K_u$-damaged (solid circles), and all-damaged (solid pentagons) cases, show a reduced switching current density (Fig. 2(b)). Most significant reduction of the switching current density is obtained for the $K_u$-damaged and all-damaged cases (Fig. 2(b)). As a result, the STT-efficiencies (= $\Delta/I_{SW}$) of these two cases are larger than that of the undamaged case (Fig. 2(c)). Especially, the all-damaged case shows about three times larger STT-efficiency than the undamaged case at the cell size of 25 nm. It suggests that a proper control of magnetic properties at the cell edge enhances the STT efficiency, which is beneficial for the performance of STT-MRAMs.

The enhanced STT efficiency of several edge-damaged cases is caused mostly by a reduced switching current density. In order to understand the origin of such reduced switching current, we focus on the $K_u$-damage as it is common in the three edge-damaged cases exhibiting a smaller switching current than the undamaged case. One may naively argue that the $K_u$-damage decreases the average effective perpendicular anisotropy <$K_{u,eff}$> of the whole cell, which is in turn responsible for the large reduction of the switching current. We note however that this average effect explains only the reduction of the thermal stability factor, not the reduction of the switching current. As shown in Fig. 2(a), the thermal stability factor (cell size = 25 nm) of the undamaged case is 46.8 whereas that of the all-damaged case is 29.6. Therefore, the thermal stability factor is reduced by a factor of about 1.6 (= 46.8/29.6). This reduction factor is consistent with the reduced <$K_{u,eff}$> of the whole cell. On the other hand, as shown in Fig. 2(b), the switching current density (cell size = 25 nm) of the undamaged case is $6.6 \times 10^6$ A/cm$^2$, whereas that of the all-damaged case is $1.6 \times 10^6$ A/cm$^2$. It leads to the reduction of the switching current by a factor of about 4.1 (= 6.6/1.6), which is much more than expected from the reduction of the thermal stability factor. One finds a similar difference of the reduction factor between the thermal stability factor and the switching current for the $K_u$-damaged case (solid upper triangles in Fig. 2(a) and (b)). It suggests that the reduced <$K_{u,eff}$> is not the main source of the largely reduced switching current.

We next show that this large reduction of the switching current in the $K_u$-damaged cells is mainly caused by a different switching mode. Figure 3(a) shows the normalized magnetization component of the undamaged and $K_u$-damaged cells at the equilibrium state. The perpendicular component <$m_z$> is almost 1 in the undamaged cell, whereas <$m_z$> deviates from 1 at the cell edges in the $K_u$-damaged cell especially at the edge. This deviation of the magnetization from the film normal (// z) in the $K_u$-damaged cell makes the STT at the initial time stage large, because the magnitude of STT is proportional to $\hat{m} \times (\hat{m} \times \hat{z})$. This large STT localized at the cell edge (i.e., magnetization tilting is localized at the edge) in turn makes the switching mode different from the undamaged cell. As shown in Fig. 3(b), the switching mode of the undamaged cell follows the well-known procedure [17], i.e., the almost in-phase precession of all magnetizations (time stages I and II) ➔ the nucleation of a magnetic domain wall at an edge (stage III) ➔ the completion of the switching via the domain wall propagation (stages IV and V). In contrast, the magnetizations at the edge of the $K_u$-damaged cell quickly tilt from the film normal (stage II) and then switch without forming a magnetic domain wall (stages III, IV, and V). This different switching mode can be also seen in the time evolution of the magnetization components (Figs. 3(c) and (d)). The $K_u$-damaged cell shows much shorter switching time (defined in Fig. 3(c)) than the undamaged cell, which is also caused by a large STT at the edge in the initial time stage.

Based on this distinctly different switching mode, the largely reduced switching current in the $K_u$-damaged cell is understood as follows. The switching current in the undamaged cell is determined by how easy to make the magnetization escape from the energy minimum direction (// ±z). As shown above, the magnetizations at the edge of $K_u$-damaged cell initially tilt from the film normal so that they experience a large STT even at the initial time stage and undergo a significant in-plane precession. These rotating edge-magnetizations supply a strong exchange field to the center-magnetizations and make the center-magnetizations easy to escape from the energy minimum direction. This is similar to the switching mode of current-induced synchronized switching [18], where an additional in-plane magnetic layer generates an alternating in-plane magnetic field on the perpendicular layer via the magnetostatic coupling and helps the perpendicular magnetization escape from the equilibrium direction. In the present edge-damaged case, the rotating magnetic field is mediated via the exchange coupling, which is much stronger than the magnetostatic coupling, so that the effect is more efficient for the reduction of the switching current than the current-induced synchronized switching [18].

Up to now, we have focused on the magnetic damage at the cell edges assuming a uniform current distribution across the MRAM cell. However, an electrical damage at the edges is also highly probable, which induces an inhomogeneous current distribution. We show numerical results of the switching current density for such case. The electrically damaged region is assumed to be the same as the magnetically damaged region with the width of 5 nm at the rim. We assume that a uniform current flows only through the electrically undamaged region for simplicity. Figure 4 shows effect of an electrical edge-damage combined with the $K_u$-damage on the switching current density. We find that the switching current of the $K_u$-damaged cell with an electrical damage can be even smaller than that of the undamaged cell with no electrical damage. We attribute this reduced switching current of the $K_u$-damaged cell with an electrical damage to the fact that the magnetizations in the undamaged region also tilt slightly from the film normal (see Fig. 3(a); $<m_z>$ distribution in the $K_u$-damaged cell). This result means that the $K_u$-damage could be effective to enhance the STT efficiency even though an additional electrical damage is present at the cell edge.

## IV. SUMMARY

We show that the STT efficiency can be largely enhanced by a proper control of magnetic properties at the MRAM cell edge. The $K_u$-damage has a key role for the enhanced STT efficiency through unexpectedly large reduction of the switching current, caused by a different switching mode. Our results suggest that careful experimental examinations are required for the etching process, especially focusing on the modification of magnetic properties at the cell edge depending on a different type of etchant or etching methodology. Experimental estimation of the modified magnetic properties at the cell edge would be challenging, but a recent progress in the experimental tools such as the ferromagnetic resonance force microscopy [19-21] may offer a way to do that.

This work was supported by the National Research Foundation of Korea (NRF) (NRF-2013R1A2A2A01013188, 2011-0028163), and KU-KIST School Joint Research Program.

## References

[1] G. I. Meijer, Science **319**, 1625 (2008).

[2] Z. Diao *et al*., Appl. Phys. Lett. **90**, 132508 (2007).

[3] J. A. Katine and E. E. Fullerton, J. Magn. Magn. Mater. **320**, 1217 (2008).

[4] S.-C. Oh *et al*., Nature Phys. **5**, 898 (2009).

[5] S. Ikeda *et al*., Nature Mater. **9**, 721 (2010).

[6] J. Z. Sun *et al*., Phys. Rev. B **88**, 104426 (2013).

[7] L. Thomas *et al*., J. Appl. Phys. **115**, 172615 (2014).

[8] A. D. Kent and D. C. Worledge, Nature Nanotechnol. **10**, 187 (2015).

[9] S. Lai, IEDM Tech. Dig. 255 (2003).

[10] R. Waser and M. Aono, Nature Mater. **6**, 833 (2007).

[11] S.-J. Choi *et al*., Adv. Mater. **23**, 3272 (2011).

[12] M. Gajek *et al*., Appl. Phys. Lett. **100**, 132408 (2012).

[13] J. C. Read, P. M. Braganca, N. Robertson, and J. R. Childress, APL Mat. **2**, 046109 (2014).

[14] R. Dittrich *et al*., J. Magn. Magn. Mater. **250**, L12 (2002).

[15] G. D. Chaves-O'Flynn, E. Vanden-Eijnden, D. L. Stein, and A. D. Kent, J. Appl. Phys. **113**, 023912 (2013).

[16] K. Song, S. C. Lee, and K.-J. Lee, IEEE. Trans. Magn. **50**, 3400704 (2014).
10

[17] D. Ravelosona *et al.*, Phys. Rev. Lett. **96**, 186604 (2006).

[18] S. M. Seo and K.-J. Lee, Appl. Phys. Lett. **101**, 062408 (2012).

[19] Z. Zhang, P. C. Hammel, and P. E. Wigen, Appl. Phys. Lett. **68**, 2005 (1996).

[20] O. Klein, G. de Loubens, V. V. Naletov, F. Boust, T. Guillet, H. Hurdequint, A. Leksikov, A. N. Slavin, V. S. Tiberkevich, and N. Vukadinovic, Phys. Rev. B **78**, 144410 (2008).

[21] H.-J. Chia, F. Guo, L. M. Belova, and R. D. McMichael, Phys. Rev. Lett. **108**, 087206 (2012).

**Figure Captions**

FIG. 1. (Color online) A schematic diagram of a MRAM cell (free layer). The red region indicates the edge-damaged region. $L$ is the cell diameter and $t$ is the thickness.

FIG. 2. (Color online) Effect of edge-damage on (a) thermal stability factor, (b) switching current density, and (c) STT efficiency.

FIG. 3. (Color online) Different switching mode by edge-damage. (a) Equilibrium magnetization profile in the undamaged and $K_u$-damaged cells (<$m_z$> and <$m_x$> are the out-of-plane and in-plane components of the magnetization). (b) Current-induced switching mode of the undamaged and $K_u$-damaged cells. Color indicates <$m_z$>-component. Time evolution of magnetization of (c) the undamaged and (d) $K_u$-damaged cells.

FIG. 4. Effect of electrical damage on the switching current.

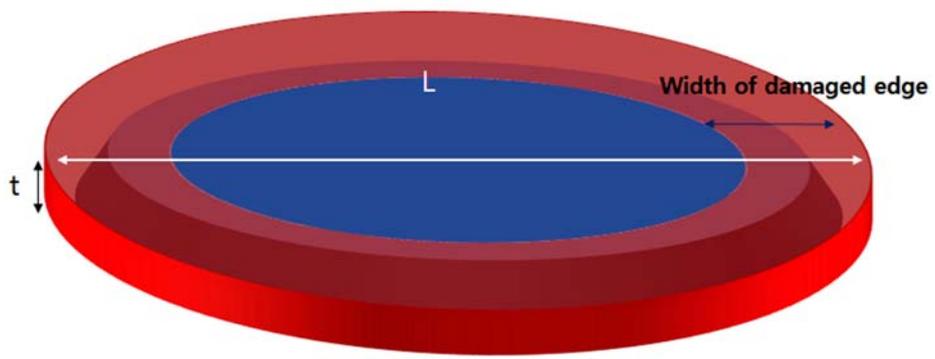

FIG. 1.

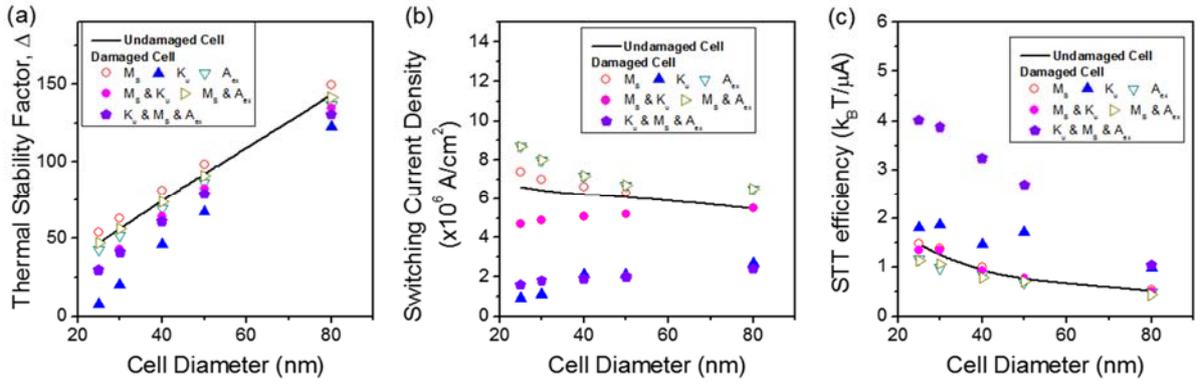

FIG. 2.

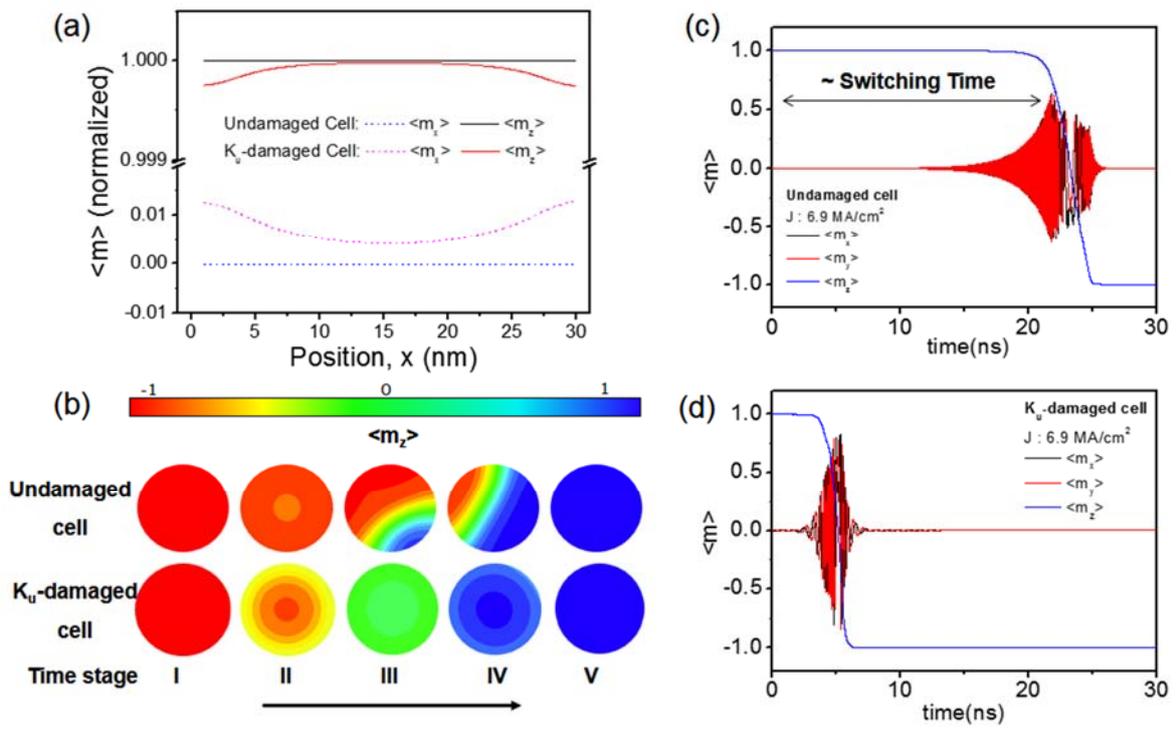

FIG. 3.

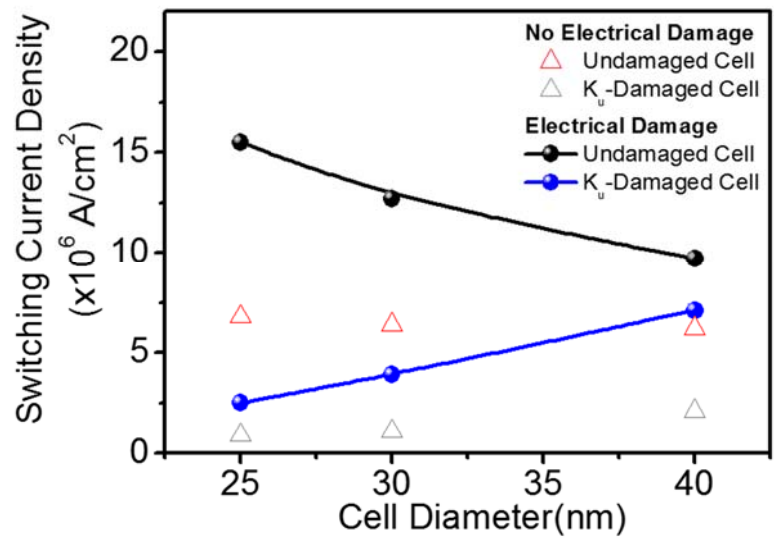

FIG. 4.